\documentclass[%
 reprint,
 amsmath,amssymb,
 aps,nofootinbib,   
]{revtex4-1}
\usepackage{bm}
\PassOptionsToPackage{linktocpage}{hyperref}
\usepackage[hyperindex,breaklinks]{hyperref}
\usepackage{enumitem}
\usepackage{slashed}

\renewcommand{\theta}{\vartheta}

\usepackage{array}
\usepackage{mathtools}

\usepackage{etoolbox}

\begin{document}

\title{Bounds on Quantum Information Storage and Retrieval} 

\author{Gia Dvali  
} 
\affiliation{%
Arnold Sommerfeld Center, Ludwig-Maximilians-University,  Munich, Germany, 
}%
 \affiliation{%
Max-Planck-Institute for Physics, Munich, Germany
}%

\date{\today}

\begin{abstract} 
We present certain universal bounds on the capacity of quantum
 information storage and on the time scale of its retrieval for 
 a generic quantum field theoretic system. 
 The capacity, quantified by the microstate entropy, is bounded from 
 above by the surface area of the object measured in units of a Goldstone decay constant. The Goldstone bosons are universally 
 present due to the spontaneous breaking of Poincare and internal
 symmetries by the information-storing object.  
  Applied to a black hole, the bound reproduces the Bekenstein-Hawking entropy. However, the relation goes beyond gravity.     
 The minimal time-scale required for retrieving the quantum information 
 from a system is equal to its volume measured in 
 units of the same Goldstone scale.  
 For a black hole this reproduces the Page time 
 as well as the quantum break-time. The same expression for 
 the information retrieval time is shared by non-gravitational  saturated states  in gauge theories, including QCD.  The saturated objects exhibit  some universal signatures such as the emission of ultra-soft radiation.  Similar bounds apply to non-relativistic many-body systems. 

  \end{abstract}

\maketitle

  The purpose of this discussion is to summarize certain universal 
  bounds on information storage and retrieval derived in 
  \cite{Dvali:2020wqi}, with preliminary concepts introduced 
  in \cite{Dvali:2019ulr, Dvali:2019jjw}.  
    
  The idea is that field theoretic objects that have a maximal 
capacity of information storage, exhibit certain universal 
properties,  insensitive to their underlying 
structure.   These properties can be quantified in the language 
of the Goldstone phenomenon \cite{Dvali:2019ulr, Dvali:2020wqi}.  

Any field configuration that is capable of storing quantum information,
breaks spontaneously 
a set of global symmetries.  These always include the Poincare symmetry. In addition, there are broken 
internal symmetries that 
parameterize various embeddings and orientations 
of the object in the internal symmetry spaces, such as the flavor or color 
spaces.  

Our focus is on self-sustained field theoretic states that have enhanced capacity of information-storage. This implies that they can exist in a 
large number $n_{st}$ of (nearly) degenerate microstates.  
 In such a case, a well-defined measure of the information storage capacity is the microstate entropy, $S \equiv \ln(n_{st})$. 

 Generically, such a ``device" represents  a macroscopic object with many elementary constituents. We shall denote their number by 
 $N$.   Let the strength of the quantum coupling among these 
 constituents be $\alpha$
and the size of the object be $R$.  For definiteness, we shall assume an approximate spherical symmetry.  Also, we shall ignore unimportant numerical factors.

  The fact that the object is self-sustained, implies that there 
  is a certain balance among $N$ and $\alpha$ at the  scale $R$. 
  If a characteristic wavelength of the constituents 
  is $R$, the self-sustainability condition reads,
  \begin{equation} \label{NA} 
  N\alpha = 1\,.   
  \end{equation} 
  At large $N$, this is the case for many 
field theoretic object such as black holes, solitons or baryons.  In more general cases, there can exist 
several sorts of constituents with different coupling. 
The bound formulated in terms of the Goldstone phenomenon is 
largely insensitive to this diversity.  
  
 In the limit of large $N$, 
 the concept of spontaneous breaking is well-defined. 
 Correspondingly, there exist gapless Goldstone modes. 
 The decay constant of a Goldstone 
  mode shall be denoted by $f$.  For a lump formed by $N$ particles of wavelengths $R$, the decay constant of the Goldstone of 
Poincare symmetry is, 
  \begin{equation} \label{fP} 
      f  =   \frac{\sqrt{N}}{R} \,.  
   \end{equation}
  The information storage capacity of the system is determined 
  by the number of distinct Goldstone ``flavors". Of course, 
  by Goldstone theorem, this is equal to a number of broken generators of the full global symmetry,
  including space-time as well as internal transformations.  
  The maximal capacity of information storage is reached when this number becomes of order $N$.  
  The different microstates of the system are the states in which 
the different sets of Goldstone flavors are excited. 
  The number of such microstates is exponential in $N$.  \\

 {\bf Bound on Information Capacity.} 
 In $4$-dimensions, the first bound on the information 
  storage capacity reads,
   \begin{equation} \label{AG} 
       S_{max}  =  {\rm Area}  \cdot f^2 \,, 
   \end{equation} 
 where ${\rm Area} \sim R^2$ is the surface area of the object. 
 Correspondingly, in  $d$ space-time dimensions we have, 
    \begin{equation} \label{AGd} 
       S_{max}  =  {\rm Area}  \cdot f^{d-2} \,, 
   \end{equation} 
where ${\rm Area} \sim R^{d-2}$.

    The second bound on entropy reads,
   \begin{equation} \label{Alpha} 
       S_{max}  =   \frac{1}{\alpha} \,,  
   \end{equation} 
   where $\alpha$ denotes the (dimensionless) running coupling among the elementary constituents, evaluated at the scale $R$ (i.e., for the momentum-transfer
 $\sim 1/R$).  The two bounds are not independent.  
 In fact, for a self-sustained system (\ref{NA}),
   they are equivalent. For example,  in $d=4$ a self-sustained 
  state of $N=1/\alpha$ particles of wavelengths $R$, breaks the Poincare 
symmetry spontaneously.  The decay constant of the corresponding 
Goldstone mode is given by (\ref{fP}).  Taking into account 
(\ref{NA}),  the expressions 
(\ref{AG}),  (\ref{CG}) and (\ref{Alpha})  are equal. \\

 Notice, the Goldstone 
 interaction strength at momentum-transfer $1/R$ is 
 $\alpha_{Gold} \equiv 1/({\rm Area}  \cdot f^{d-2})$. Thus, 
 in arbitrary dimensions,  the bound (\ref{AGd}) can also be written as, 
  \begin{equation} \label{CG} 
       S_{max}  =  \frac{1}{\alpha_{Gold}}\,. 
   \end{equation} 
 That is, for a self-sustained system (\ref{NA}) the equal bounds 
 on information capacity are set by the coupling of the collective 
 Goldstone mode of Poincare symmetry and by the coupling of the elementary constituents.

  Both bounds (\ref{AG})  and (\ref{Alpha}) are imposed by unitarity. It is evident that their violation
   would imply a violation of unitarity by  scattering 
  amplitudes.  This statement is non-perturbative and cannot  
  be cured by a re-summation.  We shall not present the detailed arguments
  \cite{Dvali:2020wqi} but shall only note that near saturation 
  the cross section of a scattering process, in units of a saturated black disk,  scales as, 
    \begin{equation} \label{Cross} 
       \sigma  =   {\rm e}^{-\frac{1}{\alpha} + S}\,.  
   \end{equation} 
  Therefore, a violation of the bound (\ref{Alpha}) would imply the violation 
  of unitarity. This is of course impossible.  A consistent theory 
  must self-prevent from entering the regime of excessive 
  entropy.  For example, there are indications \cite{Dvali:2020wqi} that the confinement
  in QCD with $N$ colors, can be 
  viewed as a mechanism preventing formation of states with 
  a color entropy exceeding the bounds (\ref{AG}) and 
  (\ref{Alpha}).

  Let us note that for a black hole the Goldstone comes from the 
  graviton. 
 Therefore, $f=M_P$, whereas 
  $\alpha = 1/({\rm Area} \cdot M_P^2)$. So, the expressions 
  (\ref{AG}), (\ref{CG}) and (\ref{Alpha}) reproduce 
 the Bekenstein-Hawking entropy \cite{BekE, Hawking}.  
 The Goldstone meaning of this 
 entropy becomes especially transparent \cite{Dvali:2015ywa} 
 in the description of a black hole as of a saturated  bound-state of  $N$ 
 soft gravitons \cite{Dvali:2011aa}.      
  However, it is evident that gravity is not special. 
 Any object of maximal entropy  (\ref{AG}) and (\ref{Alpha}) 
 satisfies the same relations.  In \cite{Dvali:2020wqi}, 
 such objects were universally referred to as ``saturons".  

Curiously, the bounds (\ref{AG}) and (\ref{Alpha}) are in general more 
stringent than the classic bound on entropy by Bekenstein
\cite{BekBound}, 
 \begin{equation} \label{BEK} 
       S_{Bek}  \sim  ER \,. 
   \end{equation} 
   It appears that (\ref{AG}) and (\ref{Alpha}) can severely limit the information capacity of the system 
   even when its entropy is well below the Bekenstein bound.
   Note, a black hole saturates all three bounds simultaneously. 
    \\

 {\bf Time of Information-Retrieval.}  
    We shall now  move to the bound on the time-scale of information
    retrieval \cite{Dvali:2020wqi}.  In terms of a Goldstone decay 
    constant, the minimal time required for the information retrieval
    in $d$-dimensions is, 
    \begin{equation}\label{tG}
     t_{min} =  {\rm Volume} \cdot f^{d-2} \,,  
     \end{equation} 
 where ${\rm Volume} \sim R^{d-1}$. 
   The alternative expression for $t_{min}$ is, 
   \begin{equation}\label{tA}
     t_{min} =   \frac{R}{\alpha}  \,. 
     \end{equation}  
  The physical meaning of the above expressions can be understood in several ways.  For example, let us use the Goldstone language. 
  As said, in this language, the information is stored in the microstates that 
    represent different states of the Goldstone modes. 
    In order to retrieve this information, the state of the Goldstone modes 
    must be detected. This requires an interaction with these modes. 
    The maximal interaction rate is, 
    \begin{equation}\label{GG}
      \Gamma_{Gold} = \frac{1}{R^{d-1}f^{d-2}} \,.
     \end{equation} 
  Correspondingly, the minimal time required for the information retrieval is inverse of this,  which gives (\ref{tG}).  
  Thus, the time is proportional to the volume. 
  Notice, (\ref{tG}) and (\ref{tA}) are absolute lower bounds. 
  The actual time-scale required for the information read-out 
  can be much longer. 
  
     Taking into account (\ref{AG}) and (\ref{Alpha}) we can 
   rewrite  (\ref{tG}) and (\ref{tA}) as, 
    \begin{equation}\label{tS}
     t_{min} =   R S  \,. 
     \end{equation}  
  As already noted in \cite{Dvali:2020wqi}, the expressions  (\ref{tA}) and (\ref{tS})  coincide with so-called quantum break-time of a generic saturated system \cite{Dvali:2017eba}, originally derived for a black hole 
  and for a de Sitter space in \cite{Dvali:2013eja}. 
  
  The physical meaning of the quantum break-time is a complete departure of the quantum evolution from the semi-classical one. It makes a lot of sense 
  that the same time-scale comes out to be the minimal time 
  required for 
  the extraction of the quantum information. This is because the extraction process 
  is intrinsically-quantum and has no classical counterpart. \\
             
  For a black hole (again, taking into account $f=M_P$ and 
  $\alpha = 1/({\rm Area} \cdot M_P^2)$),
  both (\ref{tG}) and (\ref{tA}) reproduce the so-called Page time
  \cite{Page:1993wv}. 
   As pointed out earlier,  this time  coincides with  several other 
   time-scales:  1) The quantum break-time
   \cite{Dvali:2013eja};  2) the time required for resolving 
   the black hole $1/N$ ``hair"  \cite{Dvali:2012rt};   and 
  3) the time required for resolving the departures from 
  thermality \cite{Dvali:2015aja}.  The corpuscular structure of 
  a black hole \cite{Dvali:2011aa}, explains these coincidences.   
   
  However, again, gravity appears not to be special in this respect. The same time-scale is 
  required for an arbitrary saturated system for giving away the stored information.  Several explicit examples can be found in \cite{Dvali:2020wqi}.
  
  One way of interpreting the equations (\ref{tG}), (\ref{tA}) and (\ref{tS}), is in terms 
 of explicit breaking of the vacuum degeneracy by the corpuscular $1/N$-effects. 
That is,  $1/N$-corrections lift the degeneracy and generate the energy  gaps  for the would-be Goldstone modes.
\footnote{This is somewhat reminiscent to the
 generation of the Goldstone mass-gap in $SU(N)$ QCD by $1/N$-effects via  Witten-Veneziano mechanism \cite{Witten:1979vv}, although the analogy should not be taken literally, as  
in the present case $N$ is the actual occupation number of quanta in the state. }  
  
 In general, due to the decay,  the system moves away 
 from the critical point (\ref{NA}). Due to this, the gaps of the 
 collective modes grow.    
   It takes time $t_{min}$ for 
 the gaps to grow to $\sim 1/R$ \cite{Dvali:2018xpy}.  At this point,
 the quantum state of the Goldstone modes can be resolved by a scattering experiment at the scale $1/R$.

   Notice that the bounds (\ref{tG}) and (\ref{tA}) are independent 
   of the stability of the system.  Even if the object is 
   macroscopically-stable, e.g., 
   due to a conserved topological charge, the minimal time-scale 
   for retrieving the quantum information carried by it, is still 
   given by (\ref{tG}) and (\ref{tA}).  \\
   
   {\bf Example of a Baryon.}
   We shall discuss an illustrative example   
   of the system 
   that saturates the above bounds, originally given in 
   \cite{Dvali:2019jjw, Dvali:2020wqi}.  This system is a baryon
    in $SU(N)$ QCD with $N$ colors and $N_f$ massless quark flavors.  
    Its properties are rather well understood \cite{WittenN}. 
    In particular, at large $N$ \cite{planar}
   the description of a baryon as the soliton of pions (skyrmion \cite{skyrme}) becomes solid \cite{WittenS}.      
   
  Now, a new angle \cite{Dvali:2019jjw} is to view a baryon as 
 a quantum information storing device. The quantum information is 
  stored in the flavor content of the baryon and can be quantified in terms of quarks or in terms of pions.  The number of microstates 
 $n_{st}$ is given by  the dimensionality of the representation of the flavor group under which the 
  baryon transforms.   We can bring the information capacity 
  to its maximum, by 
 increasing the number of massless quark flavors to $N_f \sim N$.       

    The maximal entropy of the baryon imposed by unitarity is, 
      \begin{equation}\label{SB}
     S_{max} =   {\rm Area}_B\cdot f_{\pi}^2  \,, 
     \end{equation}  
where ${\rm Area}_B \sim \Lambda_{QCD}^{-2}$ is the surface area of the baryon and  $f_{\pi} = \sqrt{N} \Lambda_{QCD}$ is the pion decay constant.  

 There are strong physical arguments suggesting   
that the violation of the bound  (\ref{SB}) is not possible.  First, this would render the theory asymptotically not free
due to the increase of the flavor symmetry relative to color. 
Secondly,  the scattering of pions 
  in the low energy effective theory, saturates non-perturbative unitarity
  at the saturation point of this bound.

Now, the 
minimal time-scale for reading out the information stored in flavor
quantum numbers of the baryon is \cite{Dvali:2020wqi}, 
    \begin{equation}\label{SB}
     t_{min} =   {\rm Volume}_B \cdot f_{\pi}^2  \,, 
     \end{equation}  
where ${\rm Volume}_B \sim \Lambda_{QCD}^{-3}$ is the volume 
of the baryon. \\

 For example, a baryon in a hypothetical QCD with 
 $\Lambda_{QCD} \sim 10^{-4}$eV and the number of colors and quark flavors  $\sim 10^{66}$, would have the same information
 processing characteristics as an earth mass black hole in ordinary gravity.   \\

{\bf Non-Relativistic Limit.}
Interestingly, the bounds (\ref{AG}) and (\ref{Alpha})  
have non-relativistic counterparts. These can be used to limit the 
information 
capacity of a non-relativistic system. 
Although such systems do not exhibit the full Poincare symmetry, the 
Goldstone bosons of spontaneously broken space-time translations, as well as of internal symmetries,  
do exist at large $N$. 
Correspondingly, the analog of the scale 
$f$ is well-defined.  

Likewise, 
one can always define a
non-relativistic analog of dimensionless coupling 
$\alpha$, which controls the strength of the interaction among the constituents in the Hamiltonian.  
Correspondingly, the bounds (\ref{AG}) and (\ref{Alpha})
 are well defined. 
  
What happens when we attempt to violate these bounds is that the quantum break-time of the system shortens to one oscillation period.  
Putting it differently, Gross-Pitaevskii approximation breaks down almost 
instantly.  This means that no classical macro-state can be 
defined. 

  Thus, the bounds on information storage capacity 
  (\ref{AG}) and (\ref{Alpha})
  also apply to non-relativistic systems with many degrees of freedom, 
  such as Bose-Einstein condensates of cold atoms, or even 
  quantum neural-type networks \cite{Dvali:2018vvx}.
   This explains why, around the saturation 
  points of the bounds  (\ref{AG}) and (\ref{Alpha}), such systems were 
  observed to exhibit properties strikingly similar to black holes.  
  \\
  
  {\bf Ultra-Soft Radiation.} 
 Among the universal properties exhibited by the 
objects saturating the bounds (\ref{AG}) and (\ref{Alpha}), is 
the emission of an ultra-soft radiation. 
 This term refers to a radiation with a frequency that is by $1/N$ 
 lower than the saturation 
scale $1/R$.  Putting it differently, this is the frequency,
\begin{equation} \label{Usoft}
Q_{\rm u-soft} \sim \frac{1}{t_{min}} \sim \frac{1}{NR} \sim \frac{\alpha}{R} \,,
\end{equation} 
corresponding to the information retrieval time, 
$t_{min}$, given by the equations  
(\ref{tG}),(\ref{tA}) and (\ref{tS}).

This ultra-soft radiation comes from the de-excitations of 
information-carrying Goldstone modes.  It is therefore 
an inseparable part of the saturation.  The ultra-soft radiation accompanies 
the ``harder" radiation of quanta of momenta $1/R$, which is the main 
channel of the energy loss at initial stages of  the decay. 
Its detection can provide a strong evidence 
that the object we are dealing with is a saturon.

The slow time evolution of ultra-soft Goldstone modes of frequencies (\ref{Usoft}) has been shown, both analytically and numerically, for various critical Bose-Einstein condensates 
\cite{Dvali:2015ywa, Dvali:2015wca, Dvali:2018vvx, Dvali:2018xpy}
which served as prototypes for the black hole portrait  \cite{Dvali:2011aa}.

It has been argued recently \cite{Dvali:2021ooc}
 that the color glass condensate 
in QCD \cite{Gelis:2010nm} saturates the bounds (\ref{AG}) and (\ref{Alpha}).   This, in particular, served as the basis for
establishing a 
correspondence between the color glass condensate 
of gluons \cite{Gelis:2010nm} and a description of a black hole
\cite{Dvali:2011aa} in form of a saturated condensate of gravitons. 

The wavelengths of gluons $R_S$ filling up the color glass condensate 
are considerably shorter than the QCD length. Correspondingly, 
the QCD coupling $\alpha_S$ is weak. This allows to have a 
high occupation number $N =1/\alpha_S$. 
 At initial stages, the condensate mostly decays into the quanta 
of momenta $Q_S=1/R_S$.

The prediction from saturation of the 
entropy bounds (\ref{AG}) and (\ref{Alpha}),  would be that the decay 
must be accompanied by the emission of radiation of much lower 
frequency (\ref{Usoft}). For the color glass condensate this will be given by 
$Q_{\rm u-soft} \sim Q_S/N \sim \alpha_S Q_S$. 

 A part of this radiation can escape in form of photons, avoiding 
 the hadronization and confinement. 
 This is because the colored Goldstones can annihilate into photons 
 via the quark loops.   
  Of course, the rate of such a process is suppressed by 
 powers of electromagnetic coupling 
 $\alpha_{EM}$ but is enhanced by $N$. 
 The prospects of experimental measurements is a subject of a separate 
 discussion and shall not be attempted here.

The analogous ultra-soft radiation, of frequency 
(\ref{Usoft}), 
is predicted to be emitted  by a black hole. 
In particular, this is explicit within the framework of \cite{Dvali:2011aa}. There,  it comes from the nearly-gapless 
Bogoliubov-Goldstone modes of the saturated graviton condensate.  

   This radiation is not a part of the ordinary Hawking spectrum and represents a correction to it.
It comes out in form of gravitons and other light species 
such as the photon. For large black holes, before their half-decay, 
 the rate of ultra-soft radiation 
 is extremely suppressed.  It is therefore dwarfed by the standard 
Hawking emission. 

The suppression of ultra-soft radiation  is one of the reasons why 
black holes, as well as other saturated objects, cannot radiate away the information until the time  (\ref{tA}) or (\ref{tG}). Note that by this time, it is expected that the gaps of the Goldstone modes grow, 
since after its half-decay the black hole in general moves away from 
the saturation point.
 This causes a strong back reaction 
from the information pattern to the further decay process.
This constitutes a so-called ``memory burden" effect \cite{Dvali:2018xpy}. 
Correspondingly, together with the growth of the Goldstone gaps, 
the frequency of the ultra-soft radiation is also expected 
to grow in time. 

 For micro black holes, that can potentially 
be created at colliders or in other high energy events, the 
$t_{min}$ is short.  Correspondingly, the signatures of the ultra-soft  
quanta can be much clearer. \\

The presented bounds quantify the limitations of information processing from 
the point of view of the  Goldstone phenomenon and unitarity. 
They illustrate that black holes operate within the same 
rules as generic non-gravitational objects 
of maximal information storage capacity. \\

 {\bf Acknowledgements} 
 
This work was supported in part by the Humboldt Foundation under Humboldt Professorship Award, by the Deutsche Forschungsgemeinschaft (DFG, German Research Foundation) under Germany's Excellence Strategy - EXC-2111 - 390814868,
and Germany's Excellence Strategy  under Excellence Cluster Origins.

\end{document}